\documentclass{Interspeech}

\interspeechcameraready

\usepackage{comment}
\usepackage{tabularx}
\usepackage{pifont}
\usepackage{subfig}
\usepackage{adjustbox}
\usepackage{multirow}
\usepackage{multicol}
\usepackage{stackengine}

\usepackage[dvipsnames, table, xcdraw]{xcolor} 
\usepackage{pifont} 
\definecolor{bsRed}{rgb}{0.95, 0.0, 0.0}
\usepackage{hyperref}
\definecolor{bsRed}{rgb}{0.95, 0.0, 0.0}
\definecolor{mygray}{RGB}{128,128,128} 
\definecolor{lightgray}{RGB}{200,200,200}
\usepackage{hyperref}
\usepackage{cite}

\begin{document}






\title{
Mitigating Proxy-to-Wild Domain Gap in Deepfake Speech
}

\author[affiliation={1}]{Xuanjun}{Chen}
\author[affiliation={2}]{Yun-Shing}{Wu}
\author[affiliation={2}]{Wei-Chung}{Lu}
\author[affiliation={3}]{Claire}{Lin}
\author[affiliation={1}]{Haibin}{Wu}
\authorbreak
\author[affiliation={1,4}]{Hung-yi}{Lee}
\author[affiliation={2}]{Jyh-Shing Roger}{Jang}

\affiliation{}{Graduate Institute of Communication Engineering}{National Taiwan University}
\affiliation{}{Graduate Institute of Networking and Multimedia}{National Taiwan University}
\affiliation{}{Department of Information Management}{National Taiwan University}
\affiliation{}{}{NTU Artificial Intelligence Center of Research Excellence (NTU AI-CoRE)}



\maketitle

\begin{abstract}
Recent neural audio codec-based speech generation (CodecFake) produces highly realistic audio, posing a  challenge to existing deepfake countermeasure models. 
While using codec resynthesized speech (CoRS) as proxy data improves performance, it often suffers from limited generalization. 
We propose Domain-Shift Feature Augmentation (DSFA), which simulates ``in-the-wild'' variations by transforming deterministic feature statistics into stochastic distributions during fine-tuning. 
To evaluate generalization, we further introduce Codec-based Speech Generation Extension Evaluation (CoSG ExtEval) dataset, a more challenging extension of the CoSG Eval (from CodecFake+) dataset, featuring 40 unseen generative models and long-form audio. 
Experimental results demonstrate that combining a post-trained SSL backbone with DSFA effectively narrows the proxy-to-wild domain gap. This approach achieves state-of-the-art performance across diverse CodecFake attacks in both CoSG Eval and CoSG ExtEval. 
\end{abstract}

\section{Introduction}

Advances in speech generation technologies have greatly improved the naturalness and controllability of synthetic speech. While these developments enable a wide range of beneficial applications, they also introduce serious security risks when misused for malicious audio deepfake attacks, such as misinformation dissemination, identity impersonation, and social manipulation. To address these threats, community-driven efforts including the ASVspoof \cite{todisco2019asvspoof, Liu_2023, Wang2024_ASVspoof5} and ADD \cite{yi2022add, yi2024add2023} challenges have fostered substantial progress in deepfake speech detection. However, the rapid evolution of speech generation paradigms continues to challenge existing countermeasures (CMs).

Recently, data-driven neural audio codecs \cite{wu2024codec, wu2024codec_slt24} have emerged as a core component in modern speech generation pipelines, enabling codec-based speech generation (CoSG) systems \cite{wu2024towards}. Unlike earlier deepfake methods that relied primarily on vocoders to synthesize waveforms from acoustic features, CoSG systems reconstruct speech from discrete codec representations, leading to a new class of fake speech with artifact characteristics fundamentally different from those considered in prior anti-spoofing benchmarks. These systems are capable of generating high-fidelity speech and even cloning unseen speakers from only a few seconds of reference audio, often producing samples that are difficult for humans to distinguish from genuine speech. We refer to the task of detecting fake speech produced by such systems as CodecFake detection. 

Existing studies \cite{wu24p_interspeech, chen2025codecfake+} demonstrated that CMs trained on conventional anti spoofing datasets exhibit poor generalization performance when faced with CodecFake attacks. To address this limitation, recent research has proposed the use of codec resynthesized speech, referred to as CoRS, which is obtained by encoding and decoding genuine utterances through neural audio codecs, as a form of proxy training data. 
CoRS speech shares reconstruction artifacts with CodecFake speech, as both originate from the decoding of discrete codec representations.  
Consequently, incorporating CoRS data into the training process has been shown to significantly enhance the performance of CodecFake detection systems. 
Despite its effectiveness, the use of resynthesized proxy data \cite{chen2025towards, chen2025codec} introduces a new challenge: CMs may overly rely on codec-specific or dataset-specific artifacts present in the CoRS training data, resulting in degraded generalization to unseen codecs or CoSG systems. 
Given the rapid development of CodecFake, overcoming the generalization bottleneck inherent in resynthesized proxy data is critical for building CodecFake detection systems.

In this paper, we address the proxy-to-wild domain gap in CodecFake detection. We first leverage a post-trained SSL backbone to establish a versatile representation space sensitive to deepfake artifacts. Building on this, we propose Domain-Shift Feature Augmentation (DSFA) to improve generalization ability by modeling domain uncertainty through batch-wise statistical perturbations. 
To rigorously test our approach, we further collect Codec-based Speech Generation Extension Evaluation (CoSG ExtEval) dataset, covering a broad spectrum of recent generation paradigms. Our experiments and visualizations demonstrate that DSFA promotes domain-invariant features, significantly enhancing detection performance against evolving real-world spoofing threats.


\begin{figure*}[t]
    \centering
    \includegraphics[width=2.1\columnwidth]{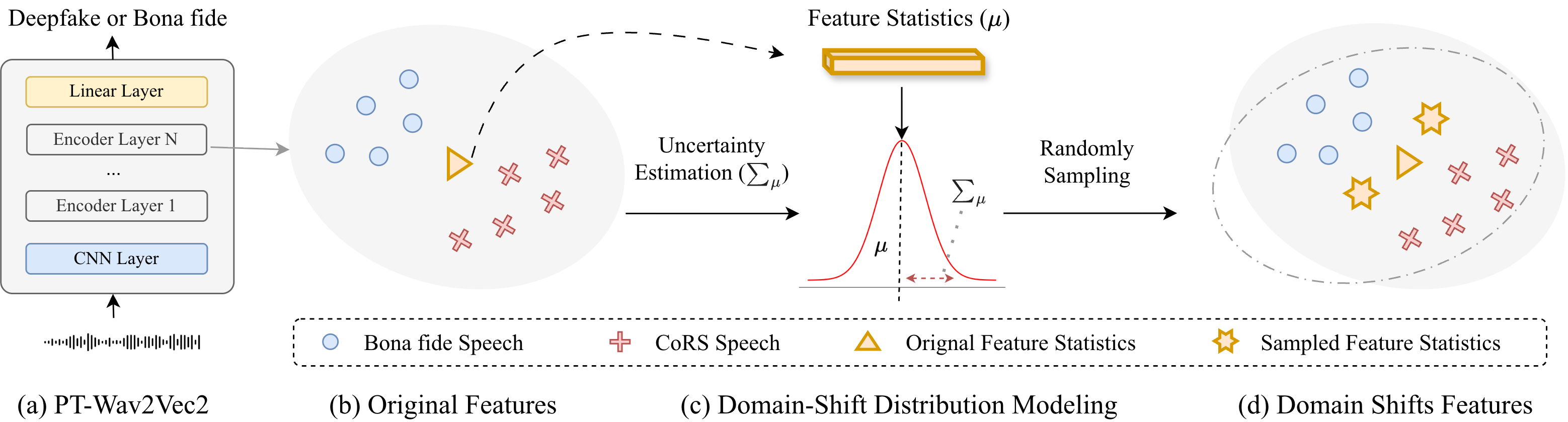}
    \vspace{-0.5em}
    \caption{Overview of the Domain Shift Feature Augmentation (DSFA) method. The proposed method estimates feature statistics $\mu$ and $\sigma$ to construct probabilistic distributions for sampling. For visual clarity, only the mean statistic $\mu$ is illustrated in this figure. }
    \label{fig:model_framework}
\end{figure*}

\section{The Proxy-to-Wild Domain Gap \\in Deepfake Speech} 
Training CMs on proxy data is a cost-effective alternative to collecting diverse TTS/VC speech \cite{wu24p_interspeech, chen2025codecfake+, wang2024can, wang2023spoofed, lu2024improving}, yet an inherent domain gap persists, hindering generalization to ``in-the-wild'' scenarios. 
We categorize this gap into three dimensions: 
\textit{(1) Artifact Mismatch:} Unseen codecs and generative models introduce unique signatures. Absent from training data \cite{chen25j_interspeech}, these novel artifacts often lead to detection failure. 
\textit{(2) Silence Mismatch:} Inconsistent pauses at utterance boundaries disrupt codec-signature alignment, causing models to overfit to background noise in silent segments rather than robust features \cite{chen2025towards, zhang2023impact}. \textit{(3) Content and Speaker Mismatch:} Unseen phonetic and prosodic patterns outside the learned distribution impair the model’s ability to distinguish bona fide from spoofed attributes \cite{chen2025towards}. 
To overcome these systemic gaps, it is essential to move beyond fixed proxy distributions and explicitly model potential domain shifts during CMs training process. 

\section{Proposed Method} Our framework (Fig.~\ref{fig:model_framework}) bridges the proxy-to-wild domain gap by: (1) leveraging a deepfake-tailored post-trained SSL backbone to establish a versatile representation space, and (2) employing Domain-Shift Feature Augmentation (DSFA) during fine-tuning to simulate unseen domain variations. 

\subsection{Post-Training Self-Supervised Learning Backbone} We initialize our model with a Self-Supervised Learning (SSL) backbone specifically post-trained for deepfake detection on a large-scale, heterogeneous corpus \cite{ge2025post} (Fig.~\ref{fig:model_framework}a). Unlike general-purpose models \cite{baevski2020wav2vec, tak2022automatic}, this backbone provides heightened sensitivity to deepfake artifacts and diverse bona fide attributes. By inheriting robust representations across varied speaker identities and codec signatures, the model establishes a versatile feature space that serves as a stable foundation for the subsequent Domain-Shift Aware fine-tuning.

\subsection{Domain-Shift Aware Fine-Tuning}
To address the performance degradation of CMs on unseen data, we propose a fine-tuning framework centered on Domain-Shift Feature Augmentation (DSFA) method. Unlike standard fine-tuning that minimizes empirical risk on fixed proxy datasets, our approach utilizes DSFA to explicitly account for potential discrepancies between proxy and target distributions (Fig.~\ref{fig:model_framework}b-d). By exposing the model to a broader range of simulated domain variations, this mechanism prevents the CMs from overfitting to proxy-specific signatures, such as those artifacts in Codec Resynthesis (CoRS) speech, thereby enhancing generalization to diverse in-the-wild scenarios. 
As illustrated in Fig.~\ref{fig:model_framework}b-c, the DSFA process facilitates this domain-aware adaptation through three primary stages: Original Feature Statistics Estimation, Domain-Shift Distribution Modeling, and Augmentation via Domain-Shift Sampling.

\textbf{Original Feature Statistics Estimation. }
To capture a compact representation of the domain ``style,'' we first extract instance-level statistics from the SSL backbone's latent feature map $x \in \mathbb{R}^{B \times C \times T}$. Specifically, we compute the channel-wise mean $\mu(x) \in \mathbb{R}^{B \times C}$ and standard deviation $\sigma(x) \in \mathbb{R}^{B \times C}$ as follows:
\begin{equation}
\begin{aligned}
    \mu_{b,c}(x) &= \mathbb{E}_t[x_{b,c,t}], \\
    \sigma^2_{b,c}(x) &=  \mathbb{E}_t \big[ (x_{b,c,t} - \mu_{b,c})^2 \big].
\end{aligned}
\end{equation}
These statistics encapsulate critical domain characteristics, such as acoustic styles and codec signatures. In our context, the domain gap between CoRS proxy data and ``in-the-wild'' CoSG samples manifests as significant shifts in these statistics. By treating $\mu$ and $\sigma$ as controllable targets, we can simulate the transition from fixed proxy distributions to unforeseen CoSG variations during training.


\textbf{Domain-Shift Distribution Modeling. }
To simulate the shift from CoRS to CoSG, we transform deterministic feature statistics into stochastic distributions. 
Drawing inspiration from latent space analysis \cite{shen2021closed, wang2019implicit}, we leverage feature variance to model the potential directions of domain shifts. 
This allows the model to simulate unseen domain variations by quantifying fluctuations within the proxy data. 
Rather than treating $\mu(x)$ and $\sigma(x)$ as fixed constants, we model their potential discrepancies as probabilistic distributions, using mini-batch variance as a proxy for domain uncertainty:
\begin{equation}
\begin{aligned}
\Sigma^2_\mu(x) &= \frac{1}{B}\sum_{i=1}^B \big(\mu(x_i)-\mathbb{E}_b[\mu(x)]\big)^2, \\
\Sigma^2_\sigma(x) &= \frac{1}{B}\sum_{i=1}^B \big(\sigma(x_i)-\mathbb{E}_b[\sigma(x)]\big)^2,
\end{aligned}
\end{equation}
where $\Sigma^2_\mu(x)$ and $\Sigma^2_\sigma(x)$ encapsulate the statistical diversity present in the current training iteration. 
These variances provide a data-driven basis for sampling perturbed domain ``styles'' that extend beyond the static boundaries of the CoRS dataset, effectively bridging the gap to unseen CoSG variations.




\textbf{Augmentation via Domain Shifts Sampling.}
To simulate potential ``in-the-wild'' variations, the DSFA module transforms deterministic feature statistics into stochastic representations via a unified perturbation scheme. Specifically, the original mean $\mu(x)$ and standard deviation $\sigma(x)$ are augmented as:
\begin{equation}
\beta(x) = \mu(x) + \epsilon_\mu \cdot \Sigma_\mu, \qquad \gamma(x) = \sigma(x) + \epsilon_\sigma \cdot \Sigma_\sigma,
\end{equation}
\begin{equation}
\epsilon =
\begin{cases}
\mathcal{U}(-1,1) , & \text{Uniform}, \\ 
\mathcal{N}(0,1) , & \text{Gaussian},
\label{lab:distribution_sel}
\end{cases}
\end{equation}
where $\epsilon$ represents stochastic noise and $\Sigma_\mu$, $\Sigma_\sigma$ modulates the perturbation magnitude. To ensure end-to-end differentiability, we employ the re-parameterization trick to decouple the randomness from the optimization process. The Uniform strategy utilizes the estimated batch-wise uncertainty $\Sigma(x)$ to bound the potential shifts, whereas the Gaussian strategy introduces standard normal noise to model the statistical fluctuations.

Following the sampling of perturbed statistics $\beta(x)$ and $\gamma(x)$, the augmented feature map is synthesized using the Adaptive Instance Normalization (AdaIN)~\cite{huang2017adain} mechanism:
\begin{equation}
\text{DSFA}(x) = \gamma(x) \left( \frac{x - \mu(x)}{\sigma(x)} \right) + \beta(x).
\end{equation}
To maintain representation stability, this augmentation is applied stochastically with probability $p$ during training:
\begin{equation}
\label{lab:indicator_function}
\hat{x} = \mathbb{I}[p_0 < p] \cdot \text{DSFA}(x) + (1 - \mathbb{I}[p_0 < p]) \cdot x,
\end{equation}
where $p_0 \sim \mathcal{U}(0,1)$ and $\mathbb{I}$ is the indicator function. This strategy encourages the model to learn representations invariant to statistical fluctuations, enhancing robustness against unpredictable domain shifts.

\textbf{Loss Function. } To enhance discriminability, we employ a joint training objective combining supervised contrastive (SupCon) \cite{khosla2020supervised} and cross-entropy (CE) losses:
\begin{equation}
\mathcal{L}_{total} = \mathcal{L}_{CE} + \lambda \cdot \mathcal{L}_{SupCon}
\end{equation}
where $\lambda$ is a balancing hyperparameter. By optimizing this joint objective, the model is encouraged to learn a more compact and discriminative embedding space, ensuring that the domain-invariant features synthesized by DSFA remain robust across diverse spoofing scenarios.

\subsection{Codec-based Speech Generation Extension Evaluation}
\label{lab:CoSG_ext}

In addition to the original CoSG evaluation set \textbf{(CoSG Eval)} in CodecFake+, we also construct an extended evaluation dataset, referred to as \textbf{CoSG ExtEval
\footnote{The \textbf{CoSG ExtEval} evaluation set and details will be released on the Github repository after the paper is accepted. }}, to further assess model generalization under more diverse and challenging conditions. 
The CoSG ExtEval dataset is collected by gathering spoofed speech samples generated from a broader set of recent codec-based speech generation models, primarily sourced from official demo pages and public repositories of state-of-the-art CoSG systems, as summarized in Table~\ref{tab:cosg_ext_info}. 
Compared to the original CoSG Eval set, CoSG ExtEval encompasses a broader spectrum of models, spanning diverse codec designs, tokenization strategies, and generation paradigms, such as autoregressive, non-autoregressive, diffusion-based, and multi-stage architectures.  

All audio samples are generated using unseen systems without any overlap with the training data. This extension is designed to better reflect real-world attack scenarios, where spoofed speech may originate from continuously evolving and previously unknown generation models, thereby providing a more rigorous benchmark for evaluating robustness and overfitting in CodecFake detection.

\section{Experimental Setup}
We conduct experiments using the CodecFake+ \cite{chen2025codecfake+} dataset, where CoRS (speech resynthesized by neural audio codecs) is employed for training and CoSG (speech from codec-based generation models) is used for evaluation. CoRS contains spoofed samples from 31 neural codecs applied to the VCTK corpus \cite{yamagishi2019cstr}. 
Following previous work \cite{chen2025codecfake+}, we adopt the taxonomy-guided balanced sampling from CoRS training dataset to select a subset of 42,965 bona fide and 42,965 spoofed samples. Specifically, we utilize the DEC balanced subset, ensuring equal representation across decoder types (time/frequency). For evaluation, we use CoSG Eval sets comprising 17 codec-based generation models sourced from their official demo pages, which reflects realistic scenarios involving unseen generative systems. 


We adopt the post-trained \textit{Wav2Vec2-Large-AntiDeepfake} model as the CM backbone \footnote{https://huggingface.co/nii-yamagishilab/xls-r-2b-anti-deepfake}. 
All inputs raw waveforms sampled at 16 kHz, cut into 4-seconds when training, with RawBoost \cite{tak2022rawboost} applied for basic data augmentation. 
Models are trained on a 3090 GPU with a batch size of 14, using the Adam optimizer with an initial learning rate of $1 \times 10^{-6}$ and weight decay of $1 \times 10^{-4}$. 
Cross-entropy loss is used for training, with the weight (0.1, 0.9) as we emphasize the bona fide samples. Performance is evaluated using Equal Error Rate (EER \%), where lower values indicate better detection performance.

\begin{table}[t]
\centering
\fontsize{7}{9}\selectfont
\setlength\tabcolsep{1.9pt}
\caption{Statistics of the CoSG evaluation datasets.}
\label{tab:cosg_ext_info}
\vspace{-1em}
\begin{tabular}{lccccc}
\toprule
\textbf{Eval Set}  &  \textbf{Types}  & \textbf{Sample} & \textbf{Models} & \textbf{DUR (h)} & \textbf{Min / Mean / Max (s)} \\
\midrule
\multirow{2}{*}{CoSG Eval.}  & Bona fide & 850  & None &  1.69 & 1.27 / 6.55 / 28.58 \\
                             & Spoofed   & 931  & 17   &  1.32 & 0.85 / 5.51 / 16.63 \\
\midrule
\multirow{2}{*}{CoSG ExtEval.} & Bona fide & 1222 & None & 1.96 & 0.81 / 5.78 / 139.67 \\
                               & Spoofed   & 1366 & 40   & 3.08 & 0.80 / 8.13 / 149.38 \\
\bottomrule
\end{tabular}
\vspace{-1em}
\end{table}

\begin{table*}[t]
\centering
\fontsize{7}{9}\selectfont 
\setlength\tabcolsep{9pt} 
\renewcommand{\arraystretch}{1.1}
\vspace{-1em}
\caption{Main Result Evaluation under CodecFake+ Dataset.}
\label{tab:Cross-Scenario}
\vspace{-1em}
\begin{tabular}{llcccccc}
\toprule
& \multirow{2}{*}{Training Data} & \multirow{2}{*}{Model Backbone} & \multirow{2}{*}{Augmentation} & \multirow{2}{*}{Loss Function} & \multicolumn{3}{c}{Testing EER (\%) $\downarrow$} \\
\cmidrule(l){6-8}
& & & & & 19LA-Eval. & CoSG Eval & CoSG ExtEval \\
\midrule
(a) & ASVspoof19 & Wav2Vec2-AASIST & RawBoost & CE loss & 0.12 & 18.92 & 32.06 \\
\midrule
(b) & CoRS (Top3) & Wav2Vec2-AASIST & RawBoost & CE loss & 1.10 & 14.09 & 38.93 \\
(c) & CoRS (Top3) + ASV19 & Wav2Vec2-AASIST & RawBoost & CE loss & 0.53 & 12.97 & 34.18 \\
(d) & CoRS (QUA Balance) & Wav2Vec2-AASIST & RawBoost & CE loss & 1.93 & 21.93 & 37.12 \\
(e) & CoRS (AUX Balance) & Wav2Vec2-AASIST & RawBoost & CE loss & 2.18 & 15.02 & 29.19 \\
(f) & CoRS (DEC Balance) & Wav2Vec2-AASIST & RawBoost & CE loss & 1.51 & 11.91 & 27.07 \\
\midrule
\rowcolor{gray!10} (g) & None & PT-Wav2Vec2 & RawBoost & CE loss & 0.11 & 3.95 & \underline{22.19} \\
\rowcolor{gray!10} (h) & CoRS (DEC Balance) & PT-Wav2Vec2-FT & RawBoost & CE loss & 0.07 & 3.56 & \underline{22.19} \\
\rowcolor{gray!10} (i) & CoRS (DEC Balance) & PT-Wav2Vec2-FT & RawBoost & CE+SupCon & 0.19 & \underline{3.00} & 24.08 \\
\rowcolor{gray!10} (j) & CoRS (DEC Balance) & PT-Wav2Vec2-FT & RawBoost+DSFA & CE+SupCon & 0.07 & \textbf{2.78} & 23.00 \\
\rowcolor{gray!10} (k) & CoRS (DEC Balance) & PT-Wav2Vec2-FT & RawBoost+DSFA & CE loss & \underline{0.08} & \underline{3.00} & \textbf{21.80} \\
\bottomrule
\addlinespace[3pt]
\multicolumn{8}{l}{\textit{\scriptsize * (a)–(f) denote dataset benchmarks; (g)–(h) are post-training SSL baselines; (i)–(k) represent our proposed methods.}}
\end{tabular}
\end{table*}

\section{Main Results}

Table \ref{tab:Cross-Scenario} presents the cross-scenario results. Beyond the benchmarks (a)–(f) from CodecFake+ \cite{chen2025codecfake+} on existing sets (ASVspoof19 LA, CoRS, CoSG Eval), we further evaluate performance on our new collected CoSG ExtEval dataset. 

\textbf{CoSG ExtEval Baseline Evaluation.} 
Model (a) achieves near-perfect in-domain results but generalizes poorly to CoSG Eval, with performance further degrading on CoSG ExtEval. Similarly, the top-tier CoRS-trained model (b) fails to improve ExtEval results. While dataset pooling (c) mitigates domain mismatch, it still underperforms relative to (a). Models (e)–(f) demonstrate that taxonomy-guided balancing is crucial for generalization; notably, the DEC-balanced strategy (f) outperforms QUA (d) and AUX (e). 
However, models exhibit worse EERs on CoSG ExtEval compared to CoSG Eval, though DEC (f) remains the most effective, consistent with prior findings \cite{chen2025codecfake+}. 

\textbf{Post-Training SSL Backbone.}
We first observe that directly adopting a post-trained SSL backbone significantly outperforms the traditional backbones in models (a)–(f), achieving EERs of 3.95\% and 22.19\% on CoSG Eval and CoSG ExtEval, respectively. Among models (h)–(j), both SupCon loss and DSFA augmentation further enhance performance on CoSG Eval, reaching EERs as low as 2.78\%. However, despite these gains, the top-performing models (i) and (j) exhibit performance degradation on CoSG ExtEval compared to the naive fine-tuning of model (h). 
This suggests that while \texttt{SupCon} loss may compromise generalization ability. 
Notably, model (k) with DSFA-only configurations demonstrate stronger robustness, achieving more consistent EER performance gains on both CoSG Eval and CoSG ExtEval. 

\textbf{Discussion of CoSG ExtEval.}
Although CoSG ExtEval uses the same collection method as CoSG Eval, it is significantly more challenging. Beyond the broader model coverage shown in Table~\ref{tab:cosg_ext_info}, we believe this performance drop stems from a fundamental gap between short and long audio. Specifically, we hypothesize that longer audio clips~\cite{Liu2025LENSDF} introduce a level of complexity that, combined with acoustic and linguistic mismatches~\cite{zang2024singfake, chen2024singing, chen2025how, huang-etal-2025-speechfake}, causes the samples in CoSG ExtEval to struggle far more than it does with the samples in CoSG Eval. We will further investigate these influencing factors in future work.

\begin{table}[t]
\centering
\fontsize{7}{8}\selectfont
\setlength\tabcolsep{4pt}
\vspace{-0.8em}
\caption{Ablation study results on layer-wise feature augmentation under different distributions, evaluated on the CoSG evaluation dataset (CoSG Eval.) and its extension (CoSG ExtEval.).}
\vspace{-1em}
\label{tab:layer-wise_analysis}
\begin{tabular}{ccc|cc}
\toprule
\multirow{4}{*}{Layer} & \multicolumn{4}{c}{Testing EER (\%) $\downarrow$} \\

\cmidrule(lr){2-5}
&  \multicolumn{2}{c}{Uniform} & \multicolumn{2}{c}{Gaussian} \\

\cmidrule(lr){2-3} \cmidrule(lr){4-5}
  & CoSG Eval & CoSG ExtEval & CoSG Eval & CoSG ExtEval \\

\midrule
Baseline  &  3.00  & 24.08  & 3.00 & 24.08  \\
\midrule
1  &  3.12  & 23.11 & \textbf{2.78} & \textbf{22.61}  \\
6  &  3.12  & 23.00 & 3.00 & 23.58  \\
12 &  3.00  & 23.50 & 3.00 & 24.00  \\
18 &  3.00  & 23.93 & 3.45 & 24.58  \\
24 & \textbf{2.78} & \textbf{22.85} & \textbf{2.78} &  23.00  \\
\bottomrule
\end{tabular}
\end{table}


\begin{table}[t]
\centering
\fontsize{7}{8}\selectfont
\setlength\tabcolsep{8pt}
\caption{Ablation study of DSFA with different probabilities $p$.}
\vspace{-1em}
\label{tab:dsfa-prob}
\begin{tabular}{l c c c c c}
\toprule
Probability ($p$) & 0.00 & 0.25 & 0.50 & 0.75 & 1.00 \\
\midrule
CoSG Eval     & 3.00 & \textbf{2.78} & \textbf{2.78} & \textbf{2.78} & \textbf{2.78} \\
CoSG ExtEval  & 24.08 & \textbf{22.77} & 23.00 & 23.08 & 24.00 \\
\bottomrule
\end{tabular}
\vspace{-1em}
\end{table}




\begin{figure}[t]
\vspace{-1.8em}
    \centering
    \subfloat[\footnotesize \textnormal{Baseline Mean}]{%
        \includegraphics[width=0.45 \linewidth]{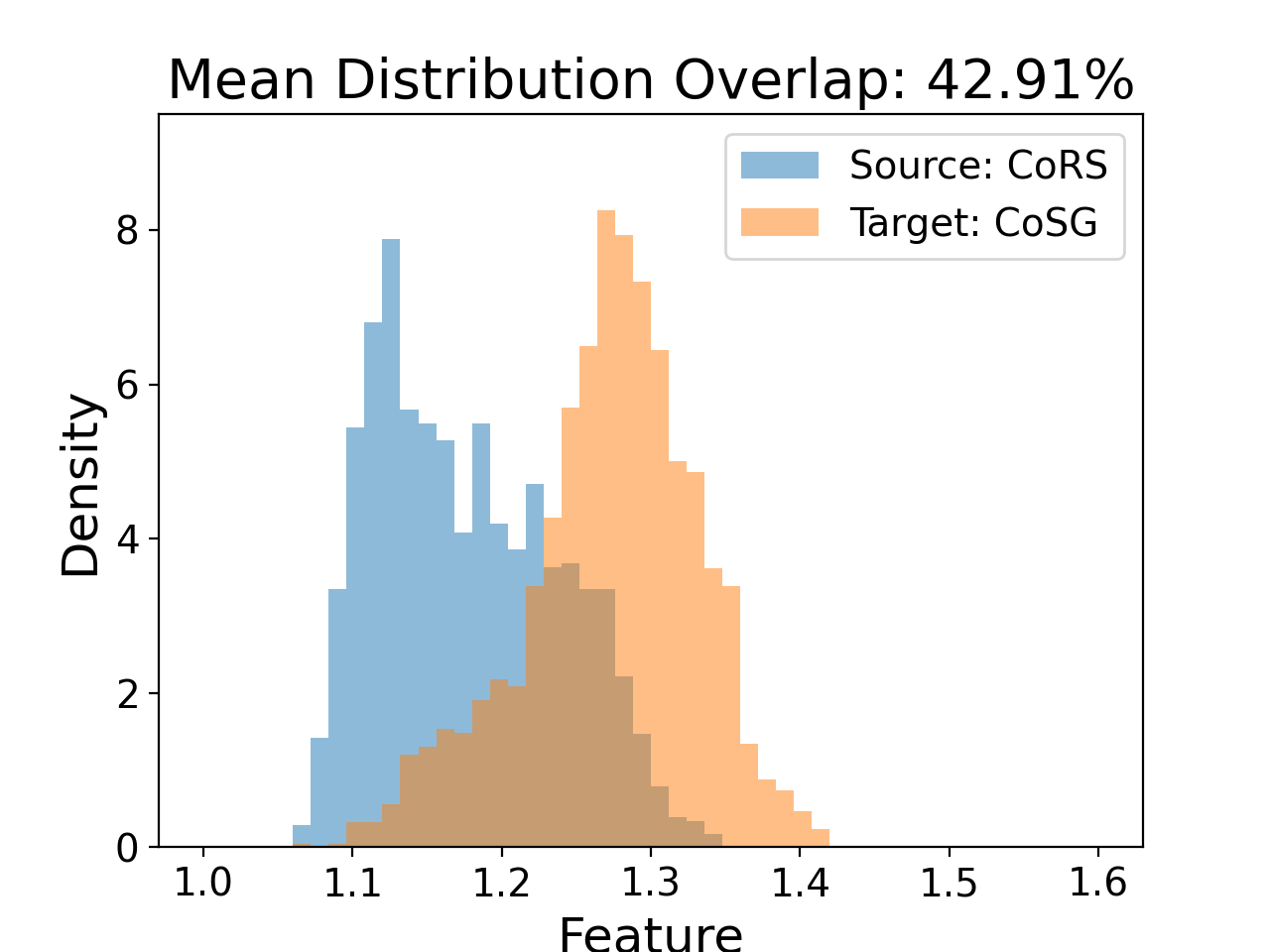}
        \label{fig:baseline}
    }
    \hfill
    \subfloat[\footnotesize \textnormal{DSFA Mean}]{%
        \includegraphics[width=0.45 \linewidth]{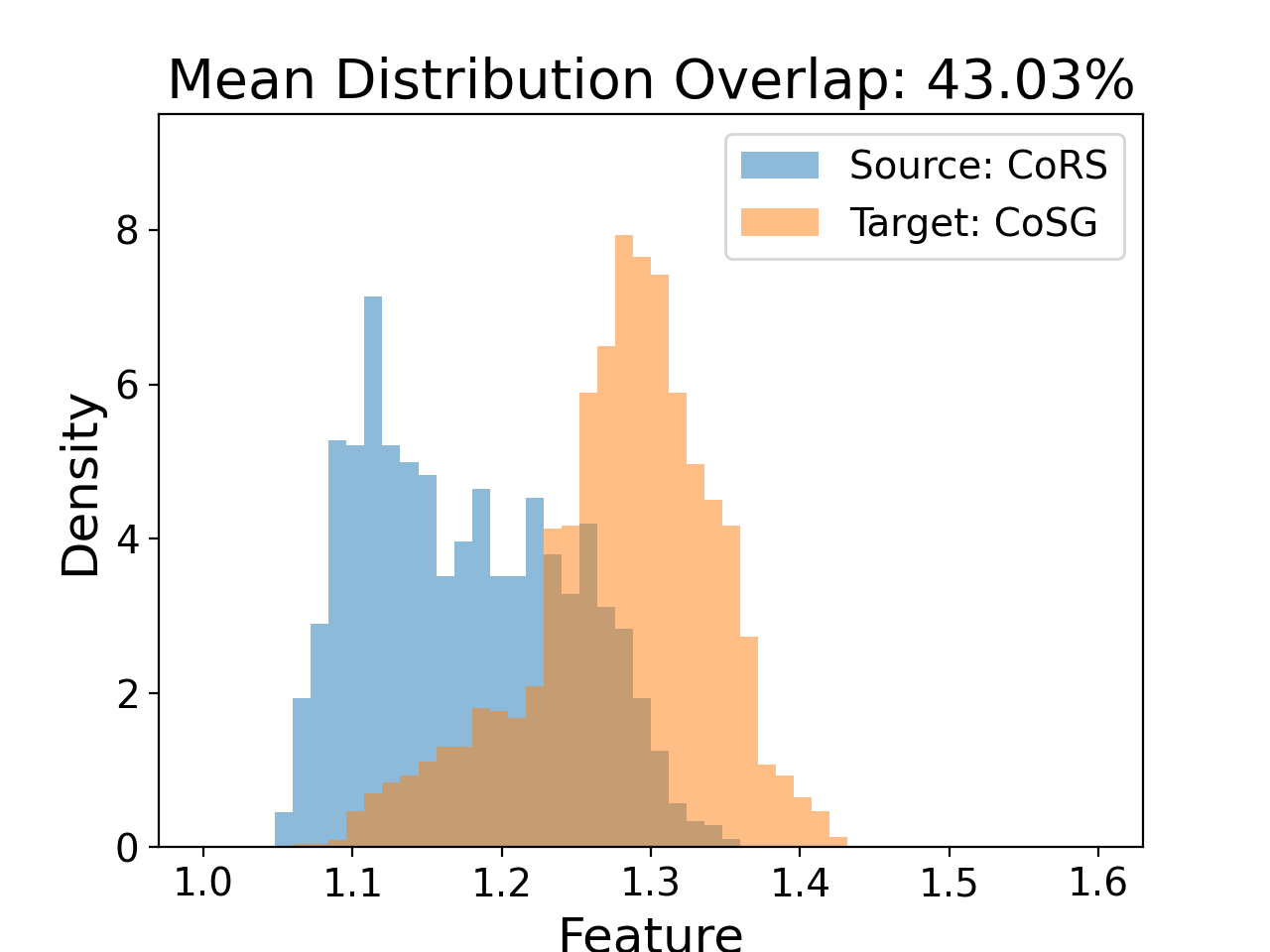}
        \label{fig:our}
    }

    \vspace{-1em}
    \subfloat[\footnotesize \textnormal{Baseline Standard Derivation}]{%
        \includegraphics[width=0.45 \linewidth]{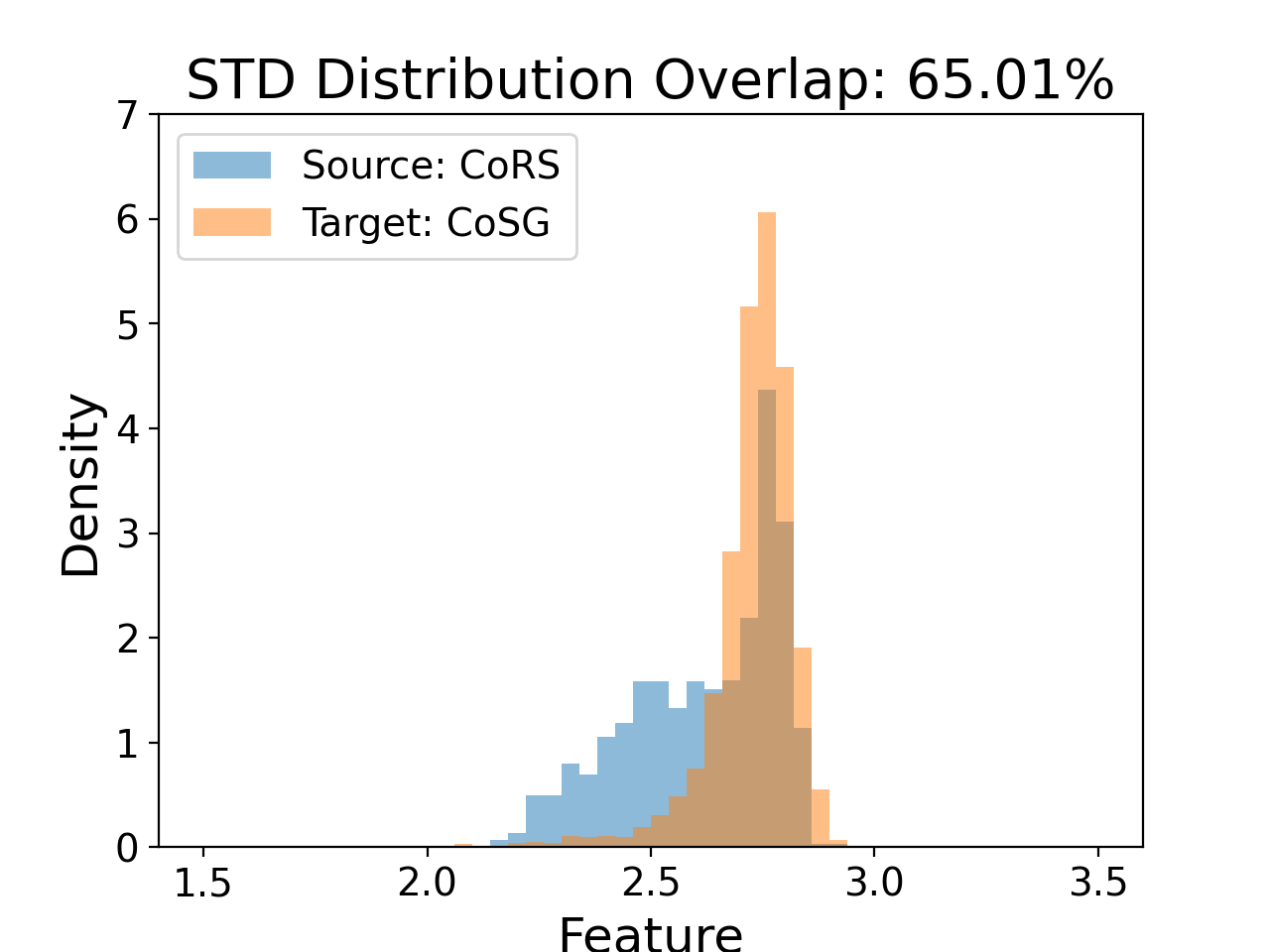}
        \label{fig:m2_vq}
    }
    \hfill
    \subfloat[\footnotesize \textnormal{DSFA Standard Derivation}]{%
        \includegraphics[width=0.45 \linewidth]{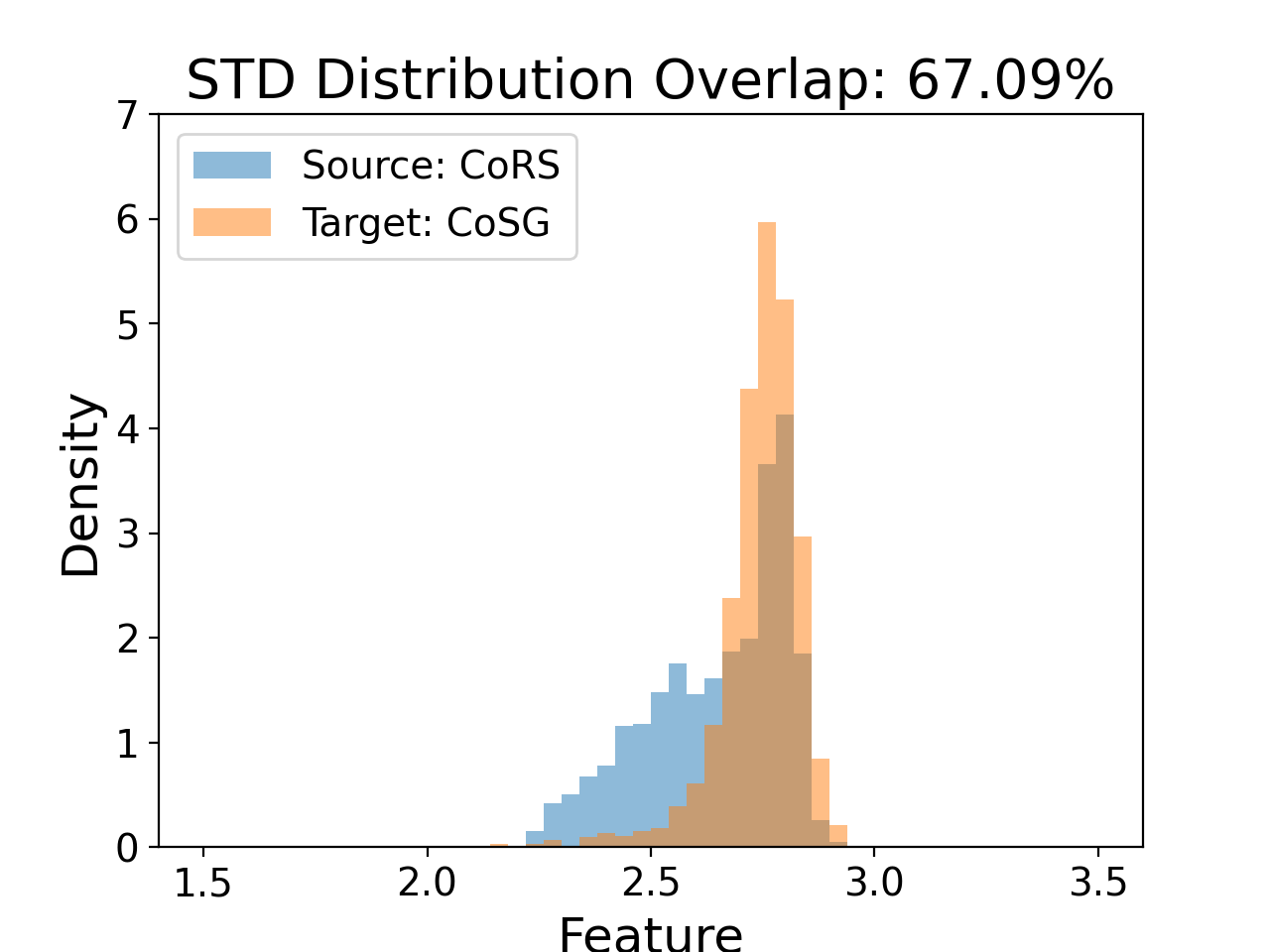}
        \label{fig:m2_aux}
    }
        \vspace{-0.5em}
    \caption{The Proxy-to-Wild Domain Gap Analysis.}
    \vspace{-3mm}
    \label{fig:distribution_analysis}
\end{figure}

\section{Ablation and Quantitative Evaluation} To further dissect the mechanisms behind these improvements and optimize feature-level augmentations, we conduct a detailed ablation study and quantitative analysis in this section.

\textbf{SSL Layer-wise Analysis. } 
We evaluate DSFA across SSL layers to identify the optimal integration point for robustness and generalizability. The impact of noise distributions (Eq. \ref{lab:distribution_sel}) is summarized in Table \ref{tab:layer-wise_analysis} (baseline: model i). Notably, the optimal layer depends on the distribution: Uniform peaks at Layer 24 (2.78\%/22.85\% EER), whereas Gaussian excels at Layer 1 (2.78\%/22.61\%) across CoSG Eval and CoSG ExtEval.

\textbf{Augmentation Ratio.}
To investigate the impact of the augmentation ratio, we evaluate different probabilities $p$ for DSFA (Eq. \ref{lab:indicator_function}), as shown in Table \ref{tab:dsfa-prob}. 
Compared to the baseline without DSFA ($p=0.00$), mostly tested probabilities improve performance on CoSG ExtEval, demonstrating the regularization benefits of DSFA. 
The optimal configuration is achieved at $p=0.25$, yielding the lowest EER of 22.77\% on CoSG ExtEval while maintaining a competitive 2.78\% on CoSG Eval. Notably, performance begins to degrade as $p$ approaches 1.00, suggesting that an excessively high augmentation rate may introduce redundant noise that hinders robust feature learning. 

\textbf{Qualitative Domain Gap Analysis.} Following \cite{li2022uncertainty_dsu}, we visualize the proxy-to-wild domain gap between CoRS (source) and CoSG (target) based on CoSG Eval in Fig. \ref{fig:distribution_analysis}. We take the intermediate features after the first encoder layer from the Transformer model and measure the feature statistics distributions of training and testing domain. While the baseline shows significant feature shifts (Figs. 2a, 2c), DSFA (Figs. 2b, 2d) improves distribution overlap for Mean ($42.91\% \rightarrow 43.03\%$) and STD ($65.01\% \rightarrow 67.09\%$). 
By narrowing the statistical gap in the latent space, DSFA aligns the data distributions and promotes domain-invariant features. This ensures that the model generalizes better from synthetic training data to real-world, in-the-wild CodecFake speech samples. 


\section{Conclusion}
This work addresses the proxy-to-wild domain gap in CodecFake detection, where models trained on resynthesized data (CoRS) exhibit a distributional bias that impairs their performance against unseen generative systems. To overcome this, we propose Domain-Shift Feature Augmentation (DSFA), which promotes domain-invariant representations by simulating statistical discrepancies in the latent space. Furthermore, we introduce CoSG ExtEval, a challenging in-the-wild evaluation set that extends the scope of CodecFake+. 
By combining a post-trained SSL backbone with DSFA for model training, our approach achieves SOTA performance on both CoSG Eval (from CodecFake+) and CoSG ExtEval, ensuring better generalization from resynthetic proxies to demanding real-world samples.  

\section{Acknowledgements}
This work was supported by the Ministry of Education (MOE) of Taiwan under the project Taiwan Centers of Excellence in Artificial Intelligence, through the NTU Artificial Intelligence Center of Research Excellence (NTU AI-CoRE).  
We thank the National Center for High-performance Computing (NCHC) of the National Applied Research Laboratories (NARLabs) in Taiwan for providing computational and storage resources. 



\section{Generative AI Use Disclosure}
We employed Gemini for grammatical paraphrasing and language polishing to improve the manuscript's clarity. 
The AI tool was utilized solely for technical editing purposes and did not contribute to the conceptualization, data analysis, or production of any significant scholarly content in this work.

\bibliographystyle{IEEEtran}
\bibliography{refs}

@inproceedings{todisco2019asvspoof,
author = {Todisco, Massimiliano and Wang, Xin and Vestman, Ville and Sahidullah, Md. and Delgado, H{\'{e}}ctor and Nautsch, Andreas and Yamagishi, Junichi and Evans, Nicholas and Kinnunen, Tomi H and Lee, Kong Aik},
booktitle = {Proc. Interspeech},
title = {{ASVspoof 2019}: future horizons in spoofed and fake audio detection}}

@article{Liu_2023,
   title={{ASVspoof 2021}: Towards Spoofed and Deepfake Speech Detection in the Wild},
   volume={31},
  journal={IEEE Transactions on Audio, Speech and Language Processing},
   author={Liu, Xuechen and Wang, Xin and others},
   year={2023}
}

@inproceedings{Wang2024_ASVspoof5,
  title = {{ASVspoof 5}: {Crowdsourced} Speech Data, Deepfakes, and Adversarial Attacks at Scale},
  booktitle = {Proc. ASVspoof Workshop},
  author = {Wang, Xin and Delgado, H{\'e}ctor and Tak, Hemlata and Jung, Jee-weon and Shim, Hye-jin and Todisco, Massimiliano and others},
  year={2024}
}

@article{yi2024add2023,
  title={{ADD 2023}: Towards Audio Deepfake Detection and Analysis in the Wild},
  author={Yi, Jiangyan and Zhang, Chu Yuan and Tao, Jianhua and Wang, Chenglong and Yan, Xinrui and Ren, Yong and Gu, Hao and Zhou, Junzuo},
  journal={arXiv preprint arXiv:2408.04967},
  year={2024}
}

@inproceedings{yi2022add,
  title={{ADD 2022}: the first audio deep synthesis detection challenge},
  author={Yi, Jiangyan and Fu, Ruibo and others},
  booktitle={Proc. ICASSP},
  year={2022}
}

@inproceedings{wu24p_interspeech,
  title     = {{CodecFake}: Enhancing Anti-Spoofing Models Against Deepfake Audios from Codec-Based Speech Synthesis Systems},
  author    = {Haibin Wu and Yuan Tseng and Hung-yi Lee},
  year      = {2024},
  booktitle = {Proc. Interspeech}

}

@article{chen2025codecfake+,
  title={{CodecFake+}: A Large-Scale Neural Audio Codec-Based Deepfake Speech Dataset},
  author={Chen, Xuanjun and Du, Jiawei and Wu, Haibin and Zhang, Lin and Lin, I and Chiu, I and Ren, Wenze and Tseng, Yuan and Tsao, Yu and Jang, Jyh-Shing Roger and Lee, Hung-yi},
  journal={arXiv preprint arXiv:2501.08238},
  year={2025}
}

@article{wu2024towards,
  title={Towards audio language modeling-an overview},
  author={Wu, Haibin and Chen, Xuanjun and Lin, Yi-Cheng and Chang, Kai-wei and Chung, Ho-Lam and Liu, Alexander H and Lee, Hung-yi},
  journal={arXiv preprint arXiv:2402.13236},
  year={2024}
}

@inproceedings{wu2024codec,
    title = "Codec-{SUPERB}: An In-Depth Analysis of Sound Codec Models",
    author = "Wu, Haibin  and
      Chung, Ho-Lam  and
      Lin, Yi-Cheng  and
      Wu, Yuan-Kuei  and
      Chen, Xuanjun  and
      Pai, Yu-Chi  and others",
    booktitle = "Findings Assoc. Comput. Linguist.",
    year = "2024"
}

@inproceedings{wu2024codec_slt24,
  author = {Wu, Haibin and Chen, Xuanjun and Lin, Yi-Cheng and Chang, Kaiwei and Du, Jiawei and Lu, Ke-Han and others},
  title = {{Codec-SUPERB@ SLT 2024}: A lightweight benchmark for neural audio codec models},
  booktitle = {Proc. IEEE Spoken Lang. Technol. Workshop},
  year = {2024}
}

@inproceedings{tak2022automatic,
  title={Automatic speaker verification spoofing and deepfake detection using wav2vec 2.0 and data augmentation},
  author={Tak, Hemlata and Todisco, Massimiliano and Wang, Xin and Jung, Jee-weon and Yamagishi, Junichi and Evans, Nicholas},
  booktitle={Proc. Odyssey Speaker Lang. Recognit. Workshop},
  year={2022}
}

@inproceedings{baevski2020wav2vec,
 author = {Baevski, Alexei and Zhou, Yuhao and Mohamed, Abdelrahman and Auli, Michael},
 booktitle = {Proc. NeurIPS},
 title = {wav2vec 2.0: A Framework for Self-Supervised Learning of Speech Representations},
 volume = {33},
 year = {2020}
}

@inproceedings{tak2022rawboost,
  title={Rawboost: A raw data boosting and augmentation method applied to automatic speaker verification anti-spoofing},
  author={Tak, Hemlata and Kamble, Madhu and Patino, Jose and Todisco, Massimiliano and Evans, Nicholas},
  booktitle={Proc. ICASSP},
  year={2022}
}

@article{yamagishi2019cstr,
  title={CSTR VCTK Corpus: English multi-speaker corpus for CSTR voice cloning toolkit (version 0.92)},
  author={Yamagishi, Junichi and Veaux, Christophe and MacDonald, Kirsten and others},
  journal={Univ. of Edinburgh, The Centre for Speech Technology Research
(CSTR)},
  year={2019}
}

@article{ge2025post,
  title={Post-training for Deepfake Speech Detection},
  author={Ge, Wanying and Wang, Xin and Liu, Xuechen and Yamagishi, Junichi},
  journal={arXiv preprint arXiv:2506.21090},
  year={2025}
}

@article{chen2025towards,
  title={Towards Generalized Source Tracing for Codec-Based Deepfake Speech},
  author={Chen, Xuanjun and Lin, I and Zhang, Lin and Wu, Haibin and Lee, Hung-yi and Jang, Jyh-Shing Roger and others},
  journal={arXiv preprint arXiv:2506.07294},
  year={2025}
}

@article{chen2025codec,
  title={Codec-based deepfake source tracing via neural audio codec taxonomy},
  author={Chen, Xuanjun and Lin, I and Zhang, Lin and Du, Jiawei and Wu, Haibin and Lee, Hung-yi and Jang, Jyh-Shing Roger and others},
  journal={arXiv preprint arXiv:2505.12994},
  year={2025}
}

@inproceedings{
li2022uncertainty_dsu,
title={Uncertainty Modeling for Out-of-Distribution Generalization},
author={Xiaotong Li and Yongxing Dai and Yixiao Ge and Jun Liu and Ying Shan and LINGYU DUAN},
booktitle={International Conference on Learning Representations},
year={2022},
url={https://openreview.net/forum?id=6HN7LHyzGgC}
}

@inproceedings{huang2017adain,
  title={Arbitrary style transfer in real-time with adaptive instance normalization},
  author={Huang, Xun and Belongie, Serge},
  booktitle={Proceedings of the IEEE international conference on computer vision},
  pages={1501--1510},
  year={2017}
}

@inproceedings{chen25j_interspeech,
  title     = {{Codec-Based Deepfake Source Tracing via Neural Audio Codec Taxonomy}},
  author    = {Xuanjun Chen and I-Ming Lin and Lin Zhang and Jiawei Du and Haibin Wu and Hung-yi Lee and Jyh-Shing Roger Jang},
  year      = {2025},
  booktitle = {{Interspeech 2025}},
  pages     = {1538--1542},
  doi       = {10.21437/Interspeech.2025-1297},
  issn      = {2958-1796},
}

@article{zhang2023impact,
  title={The impact of silence on speech anti-spoofing},
  author={Zhang, Yuxiang and Li, Zhuo and Lu, Jingze and Hua, Hua and Wang, Wenchao and Zhang, Pengyuan},
  journal={IEEE/ACM Transactions on Audio, Speech, and Language Processing},
  volume={31},
  pages={3374--3389},
  year={2023},
  publisher={IEEE}
}

@inproceedings{shen2021closed,
  title={Closed-form factorization of latent semantics in gans},
  author={Shen, Yujun and Zhou, Bolei},
  booktitle={Proceedings of the IEEE/CVF conference on computer vision and pattern recognition},
  pages={1532--1540},
  year={2021}
}

@article{wang2019implicit,
  title={Implicit semantic data augmentation for deep networks},
  author={Wang, Yulin and Pan, Xuran and Song, Shiji and Zhang, Hong and Huang, Gao and Wu, Cheng},
  journal={Advances in neural information processing systems},
  volume={32},
  year={2019}
}

@INPROCEEDINGS{wang2024can,
  author={Wang, Xin and Yamagishi, Junichi},
  booktitle={ICASSP 2024 - 2024 IEEE International Conference on Acoustics, Speech and Signal Processing (ICASSP)}, 
  title={Can Large-Scale Vocoded Spoofed Data Improve Speech Spoofing Countermeasure with a Self-Supervised Front End?}, 
  year={2024},
  pages={10311-10315},
}

@INPROCEEDINGS{wang2023spoofed,
  author={Wang, Xin and Yamagishi, Junichi},
  booktitle={ICASSP 2023 - 2023 IEEE International Conference on Acoustics, Speech and Signal Processing (ICASSP)}, 
  title={Spoofed Training Data for Speech Spoofing Countermeasure Can Be Efficiently Created Using Neural Vocoders}, 
  year={2023},
  pages={1-5}
}

@inproceedings{lu2024improving,
  title={Improving copy-synthesis anti-spoofing training method with rhythm and speaker perturbation},
  author={Lu, Jingze and Zhang, Yuxiang and Li, Zhuo and Shang, Zengqiang and Wang, Wenchao and Zhang, Pengyuan},
  booktitle={Interspeech},
  volume={2024},
  pages={512--516},
  year={2024}
}

@article{khosla2020supervised,
  title={Supervised contrastive learning},
  author={Khosla, Prannay and Teterwak, Piotr and Wang, Chen and Sarna, Aaron and Tian, Yonglong and Isola, Phillip and Maschinot, Aaron and Liu, Ce and Krishnan, Dilip},
  journal={Advances in neural information processing systems},
  volume={33},
  pages={18661--18673},
  year={2020}
}

@inproceedings{zang2024singfake,
  title={Singfake: Singing voice deepfake detection},
  author={Zang, Yongyi and Zhang, You and Heydari, Mojtaba and Duan, Zhiyao},
  booktitle={ICASSP 2024-2024 IEEE International Conference on Acoustics, Speech and Signal Processing (ICASSP)},
  pages={12156--12160},
  year={2024},
  organization={IEEE}
}

@inproceedings{chen2024singing,
    title={Singing Voice Graph Modeling for SingFake Detection},
    author={Xuanjun Chen and Haibin Wu and Jyh-Shing Roger Jang and Hung-yi Lee},
    booktitle={Interspeech 2024},
    year={2024}
}

@misc{chen2025how,
    title={How Does Instrumental Music Help SingFake Detection?},
    author={Xuanjun Chen and Chia-Yu Hu and I-Ming Lin and Yi-Cheng Lin and I-Hsiang Chiu and You Zhang and Sung-Feng Huang and Yi-Hsuan Yang and Haibin Wu and Hung-yi Lee and Jyh-Shing Roger Jang},
    year={2025},
    eprint={2509.14675},
    archivePrefix={arXiv},
    primaryClass={cs.SD}
}

@inproproceedings{Liu2025LENSDF,
  author = {Liu, Xuechen and Ge, Wanying and Wang, Xin and Yamagishi, Junichi},
  title = {LENS-DF: Deepfake Detection and Temporal Localization for Long-Form Noisy Speech},
  booktitle = {IEEE International Joint Conference on Biometrics (IJCB)},
  address = {Osaka, Japan},
  year = {2025},
}

@inproceedings{huang-etal-2025-speechfake,
    title = "{S}peech{F}ake: A Large-Scale Multilingual Speech Deepfake Dataset Incorporating Cutting-Edge Generation Methods",
    author = "Huang, Wen  and
      Gu, Yanmei  and
      Wang, Zhiming  and
      Zhu, Huijia  and
      Qian, Yanmin",
    booktitle = "Proceedings of the 63rd Annual Meeting of the Association for Computational Linguistics (Volume 1: Long Papers)",
    month = jul,
    year = "2025",
    address = "Vienna, Austria",
    publisher = "Association for Computational Linguistics",

    pages = "9985--9998",
}

\end{document}